\documentclass[11pt]{article}

\usepackage[top=2cm, bottom=2cm, left=2cm, right=2cm]{geometry}

\usepackage[noadjust]{cite}

\usepackage[cmex10]{amsmath}
\usepackage{amsbsy}

\usepackage{stackengine}
\usepackage{makecell}
\usepackage{boldline}
\setcellgapes{3pt}

\usepackage[latin2]{inputenc}
\usepackage{amsmath}
\usepackage{amsbsy}
\usepackage{bm}
\usepackage{amsthm,dcolumn,graphicx,multicol,fancyhdr}
\usepackage{latexsym}

\usepackage{algorithm}
\usepackage{algpseudocode}
\usepackage{varwidth}

\usepackage{tikz}
\usetikzlibrary{shapes, shadows, arrows}
\usepackage{url}

\usepackage{graphicx}
\usepackage{float}
\usepackage{stfloats}
\usepackage{wrapfig}

\usepackage{epstopdf}

\usepackage{amsthm,dcolumn,graphicx,multicol,fancyhdr}
\usepackage{latexsym}
\usepackage{graphicx}
\usepackage{ifthen}
\usepackage{rotating,amsfonts,color,amssymb,amsmath,amscd,array,enumerate,afterpage}

\usepackage{hyperref}
\usepackage{multirow}

\usepackage{enumitem}
\setlist{parsep=3pt,listparindent=\parindent}

\usepackage[caption=false,font=footnotesize]{subfig}

\usepackage{array}
\newcolumntype{L}[1]{>{\raggedright\let\newline\\\arraybackslash\hspace{0pt}}m{#1}}
\newcolumntype{C}[1]{>{\centering\let\newline\\\arraybackslash\hspace{0pt}}m{#1}}
\newcolumntype{R}[1]{>{\raggedleft\let\newline\\\arraybackslash\hspace{0pt}}m{#1}}

\usepackage{arydshln}

\theoremstyle{definition}

\newcommand{\babc}{\renewcommand{\labelenumi}{(\alph{enumi})}\begin{enumerate}}
\newcommand{\eabc}{\end{enumerate}}
\newcommand{\biii}{\renewcommand{\labelenumi}{(\roman{enumi})}\begin{enumerate}}
\newcommand{\eiii}{\end{enumerate}}

\newcommand{\beqn}{\begin{eqnarray*}}
\newcommand{\beq}{\begin{eqnarray}}
\newcommand{\eeqn}{\end{eqnarray*}}
\newcommand{\eeq}{\end{eqnarray}}

\newcommand{\ckboldon}[1]{#1}

\newcommand{\ckbold}[1]{%
 \ifthenelse{\isundefined{\ckboldon}}{#1}{ \textbf{#1} }
}

\makeatletter
\def\hlinewd#1{%
  \noalign{\ifnum0=`}\fi\hrule \@height #1 \futurelet
   \reserved@a\@xhline}
\makeatother

\usepackage{color, colortbl}
\usepackage{gensymb}
\definecolor{Gray}{gray}{0.9}
\definecolor{LightCyan}{rgb}{0.88,1,1}

\begin{document}
\date{}
\title{Wind Data Analysis for Assessing the Potential of Off-Grid Direct EV Charging Stations}

\author{Fuad Noman\footnote{Institute of Sustainable Energy, Universiti Tenaga National, Jalan Ikram-Uniten, 43000 Kajang, Selangor, Malaysia (e-mail: fuad.noman@uniten.edu.my)},
Ammar Al-Kahtani\textsuperscript{*}, 
Vassilios Agelidis \footnote{Department of Electrical Engineering, Technical University of Denmark, Anjer Engeluds Vej 1 Bygning 101A, 2800, Kgs. Lyngby, Denmark},
Sieh Kiong Tiong\textsuperscript{*},
Gamal Alkawsi\textsuperscript{*} \\
Janka Ekanyake\textsuperscript{*}
}

\markboth{To be Submitted for review}%
{Shell \MakeLowercase{\textit{et al.}}: Bare Demo of IEEEtran.cls for Journals}

\maketitle

\begin{abstract}
The integration of large-scale wind farms and large-scale charging stations for electric vehicles (EVs) with the electricity grids necessitate energy storage support for both technologies. Matching the energy variability of the wind farms with the demand variability of the off-grid EVs could potentially eliminate the need for expensive energy storage technologies required to stabilize the grid. The objective of this paper is to investigate the feasibility of using wind generation as direct energy source to power the EV charging stations. An interval-based approach corresponding to the time slot of EV charging is introduced for wind energy conversion and analyzed using different constrains and criteria including, wind speed averaging time interval, various turbines manufacturers, and standard high-resolution wind speed data sets. To measure the EV charging efficiency, a quasi-continuous wind turbines' output energy is performed using a piecewise recursive approach. Wind averaging results show that the three minutes intervals have increased the total number of EVs by more than 80\% compared to one-- and two -minute intervals. The potential cost reduction due to decoupling of both technologies from the utility grid, energy storage systems, and the associated energy conversion power electronics has merit and research in this direction is worth pursuing.\\
\end{abstract}

\vspace{-0.02in}

\vspace{-0.05in}

\section{Introduction} 
\label{sec:introduction}
The rapid increase of electric vehicles' (EV) adoption has set a great milestone to achieve in combating the ever increase of green house gas (GHG) emissions worldwide \cite{sun2020technology}. However, with this great goal, comes huge challenges. Charging infrastructure is expected to spike with the increasing number of EVs resulting in a huge electricity demand. Besides, several power quality issues related to grid connections and integration with these newly added nonlinear loads and high frequency switching converters are to arise \cite{das2019electric, khalid2019comprehensive}. Hence, an ascending demand of different types of  renewable energy based EV charging stations is in a race in many countries \cite{fathabadi2018utilizing}.

Although, the currently implemented EV charging stations are mostly powered by electric power distributed by utility grids \cite{dong2015phev, luo2019joint}, other charging stations designs powered by various renewable energy sources have been reported in the literature. EV charging stations powered by solar and grid's electric power are studied in \cite{goli2014pv,brenna2014urban,nunes2015day,mouli2016system, fathabadi2017novelsolarwind}. There are few studies have used the grid connected wind energy for EV charging, as in \cite{fathabadi2017novelwind, goli2014wind}. Also, other studies of hybrid stand-alone renewable sources have been reported for optimal management, control, and dispatch strategies for EV charging stations \cite{guzzi2017integration, clairand2018electric, khooban2017load, alharbi2017optimal, sanchez2019methodology}. Although these studies have proposed efficient solutions for stand-alone renewable EV charging stations, they mostly reported their theoretical results using simulation environments using limited data rather than real world data and real case studies.

The related size optimization methodologies and design problems of EV charging stations for stand-alone solar and wind microgrid was presented in \cite{al2017review, dominguez2019design} concluding that a mix of renewable energy sources and storage system is best for optimal design of EV charging stations. With many studies focused in using less variable renewable sources, i.e. solar power, a large energy demand of EV  charging stations will require larger areas covered by solar cells. Wind energy is one of the solutions to rely on for this type of charging infrastructure supported by fast charging technology. However, the inherent intermittency challenges of wind power \cite{chen2017multi} was the main barrier to develop a stand-alone wind-EV charging stations. Fast charging technology is a key to increasing the expansion of EV adoption because it removes one of the barriers that have set many consumers back from purchasing their first EVs. However, establishing fast charging points not only requires a parallel increase in the power supply from the utility companies, but also induces significant negative impacts and degrades the power quality at the grid side during the AC/DC conversions \cite{khalid2019comprehensive, zheng2019integrating}. 

In previous studies, integrating energy storage (i.e. backup batteries) into the power system was the easiest way to reduce the intermittency,
unpredictability and power fluctuations; therefore, providing a stable and continuous renewable electricity supply \cite{sanchez2019methodology}. The necessity of adding battery banks as energy storage is considered as the main defect of stand-alone charging stations. This is due to the high installation and maintenance cost, and the relatively short life span when repeatedly charged and discharges in daily basis.  Recent study by \cite{fathabadi2020novel} suggested to replace the conventional battery storage and use permanent lifetime fuel cells system as supporting power source which produce extra electric energy when renewable (i.e. solar and wind) are not available, at the same time, it produces hydrogen from the excess energy of stand-alone renewables if the harvested energy was more than EV demands. An accurate prediction of wind power and the implementation of adaptive maximum power point tracking (MPPT) algorithms are vital parts of the successful establishment of these proposals.    

In this paper, the feasibility of using fast charging technology with stand-alone wind energy source for EV charging. Unlike previous studies which implemented energy system analysis using hourly simulated distributions, we used real wind speed measurements averaged in minutely basis obtained from two different sites within two distances years. We perform extensive data analysis and benchmark comparisons using different scenarios of possible constraints on the introduced EV charging methodology. A piecewise recursive approach is implemented to study the wind data on interval-based manner to provide more precise estimates of the wind power stability over short time intervals as well as the EV charging station capacity. A case study of using fast charging supported Tesla model 3 standard range plus EV is reported and discussed in this paper. The main contributions of this work are the following: (1) we rigorously evaluate the possibility of using stand-alone wind energy source for direct fast EV charging; (2) we conducted wind speed and power analysis using real-world high sampled data sets of two years; (3) we also evaluated different wind turbines using large library (i.e. 86) of power curves; and (4) We examine different averaging time intervals for wind speed and their effect on the power output stability of wind turbines.

The rest of the paper is organized as follows. Section II presents the wind data used in this study and outlines the related data preprocessing steps. Section III describes the performance of different technologies for direct use of wind energy as a stand-alone source to power either large-scale or fast-charging EV stations. Section IV reports the analysis and comparative evaluation between the introduced methods under different criteria, as well as the verification and discussion of such results. Finally, Section V summarises the conclusions of the study.

\section{Wind Data and Preprocessing} \label{Sec:2B}
We used the publicly available database obtained from the National Renewable Energy Laboratory's National Wind Technology Center (NWTC) \cite{clifton2014135}. The NWTC is located approximately 8 km (5 miles) south of Boulder, Colorado, 36 km (20 miles) north-west of Denver (United States). We used both M2 and M4 towers data with M2 tower is located on the western end and M4 in the southwest corner of the NWTC grounds. The data from M4 tower are obtained at 20 Hz including various atmospheric properties (i.e. wind speed, wind direction. temperature, dew point, pressure, acceleration, precipitation) collected from different heights (3 to 134 m). The M2 tower data were taken every two seconds and averaged over one minute measured at different heights (from 2 to  80 m). One year data were selected for the analysis in this paper considering the data quality, 2004 and 2018 for M2 and M4 respectively.

\begin{figure}[!t]
	\centering
	\includegraphics[width=0.56\linewidth,keepaspectratio]{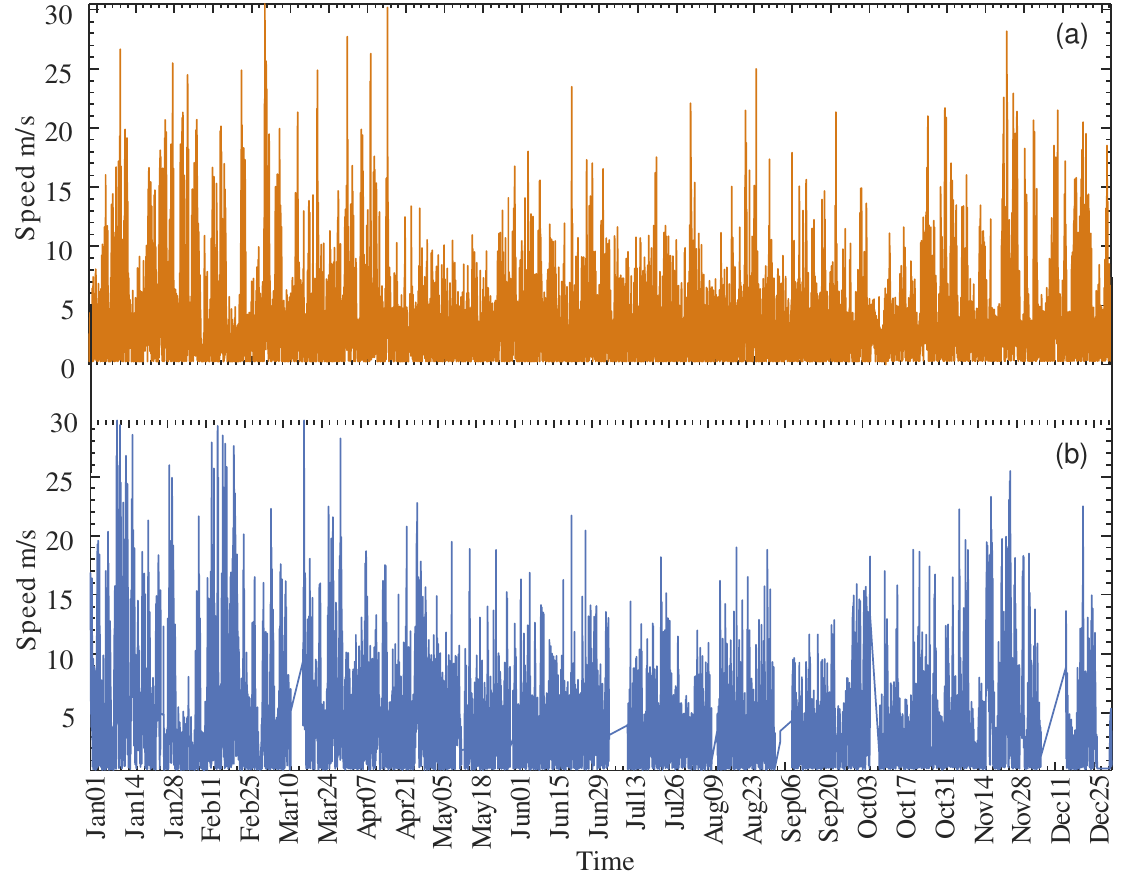}
\vspace{-0.25in}
\caption{Example of one minute wind speed data, (a) M2-Tower; (b) M4-Tower}
\label{Fig:wins_speed}
\end{figure}

The wind speed data were pre-processed in which all the NaN and encoded values are imputed using sliding window average. Let the wind speed data $y$ have a missing data at time $t$, the average of sliding window $\bf{y}$ $= [y_{t-2}, y_{t-1}, y_{t+1},y_{t+2}]$ was used for imputations considering the location (head or tail of data) and consecutive missing values. 
The high resolution (20 Hz) wind speed data were then averaged over different time intervals (i.e. one, two, and three minutes), to investigate the wind speed and power output variability. The wind meteorological organization (WMO) standard for estimating the mean wind is the 10-min average \cite{harper2010guidelines}. For smaller intervals, WMO has recommended wind speed conversion gust factor for four different land classes. The recommended gust factors are greater than 1 for the averaging intervals implemented in this study. Therefore, in this paper, the average wind speed is found by using a simple scalar average of the wind speed observations regardless of gust effects. Fig.~\ref{Fig:wins_speed} shows the one-minute average wind speed data of both M2 and M4 towers. It is worth noting that wind speed data of M4-tower has some missing observations for long periods of time (i.e. days, as in Fig.~\ref{Fig:wins_speed}(b)) with approximately 12\% of the total data, however, we visually inspected and chose the data considering the less missing and less noisy observations. 

\section{Methodology}
\begin{figure}[!t]
	\centering
	\includegraphics[width=0.55\linewidth,keepaspectratio]{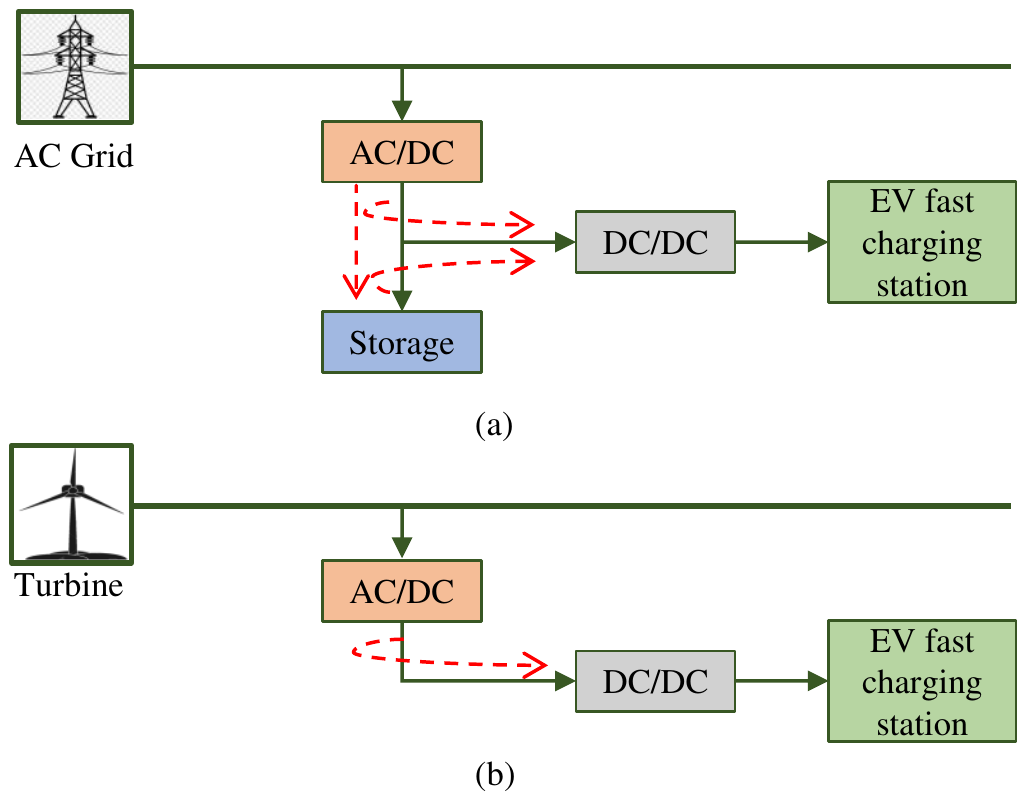}
\caption{Direct current bus configuration of fast EV charging station; (a) grid-connected with storage system; (b) direct-wind}
\label{Fig:dc_bus_config}
\end{figure}

There are two ways to utilize wind energy to charge EVs as a source. The first one is via the electricity grids where energy storage is required for both. The second could potentially be an off-grid solution in order to avoid expensive energy storage equipment, power electronics conversion equipment and conversion stages because a direct current (DC) bus system can be used in the second case. Fig.~\ref{Fig:dc_bus_config}(a) shows the AC grid interconnection of large-scale wind turbines and EV charging station. Fig.~\ref{Fig:dc_bus_config}(b) shows the off grid alternative one.

\begin{figure}[!t]
	\centering
	\includegraphics[width=0.55\linewidth,keepaspectratio]{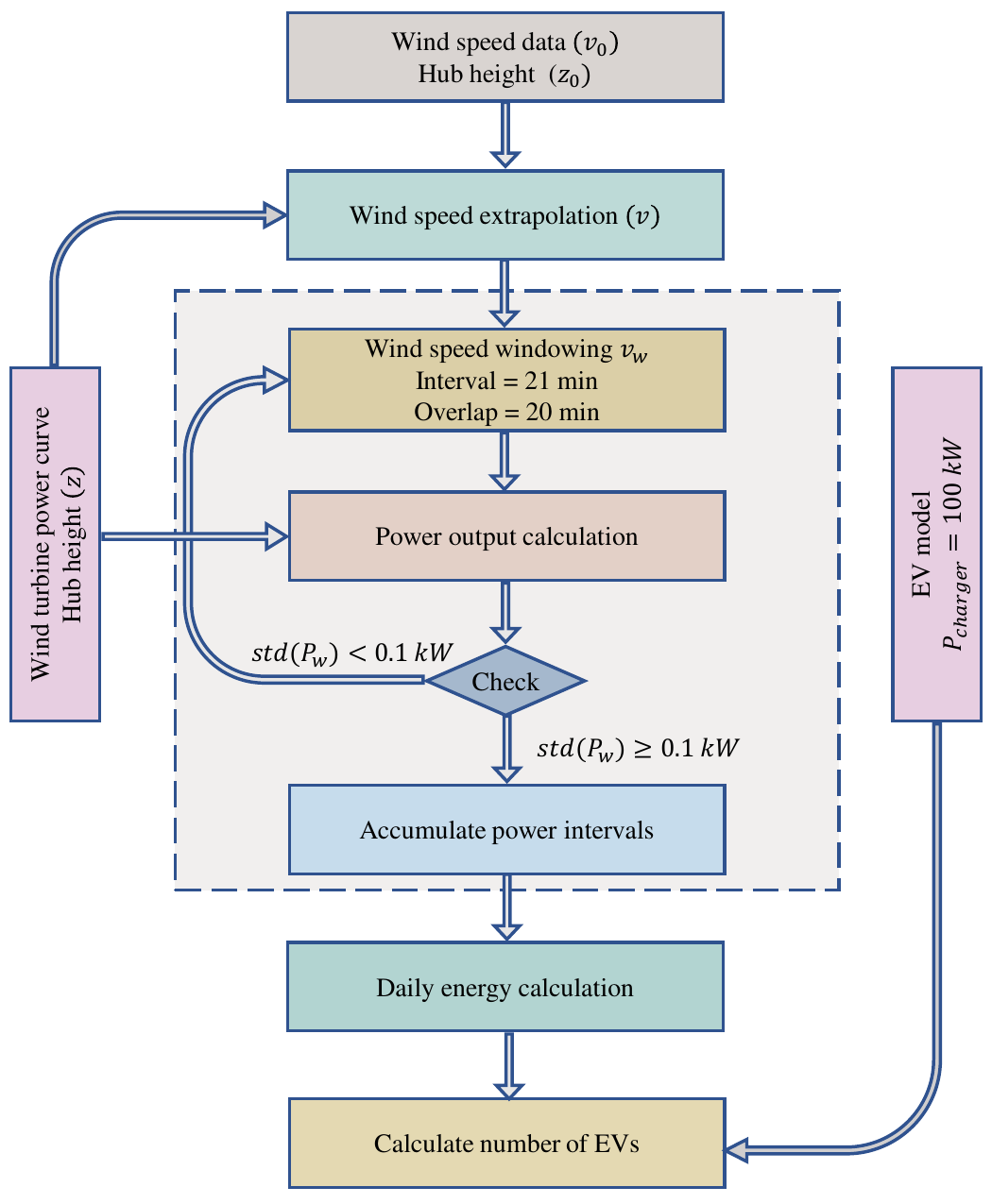}
\caption{Schematic view the data analysis procedure for off-grid wind-EV charging stations.}
\label{Fig:flow_chart}
\end{figure}

Fig.~\ref{Fig:flow_chart} shows the general procedure of isolated wind energy direct EV fast charging system. The system objective is power analysis of different wind turbines to investigate their ability to provide near-constant power supply over a specific time intervals. As shown in Fig.~\ref{Fig:flow_chart}, the wind speed data were first extrapolated to the hub height of the selected wind turbine. Then the output power of the wind turbine was calculated over overlapping time intervals. The time interval corresponding to the charging time of the EV battery (e.g. from 20\% to 80\% state-of-charge) with fast charging capability. The interval-based output power converted from the entire wind speed data were then analyzed to satisfy the near-constant (with minimum variability) power criteria. These power intervals were then used to calculate the daily energy output of the wind turbine. Using the output energy and the EV charging point characteristics, the number of EVs was calculated. 

\subsection{Wind Turbine Selection}
\begin{figure}[!t]
	\centering
	\includegraphics[width=0.55\linewidth,keepaspectratio]{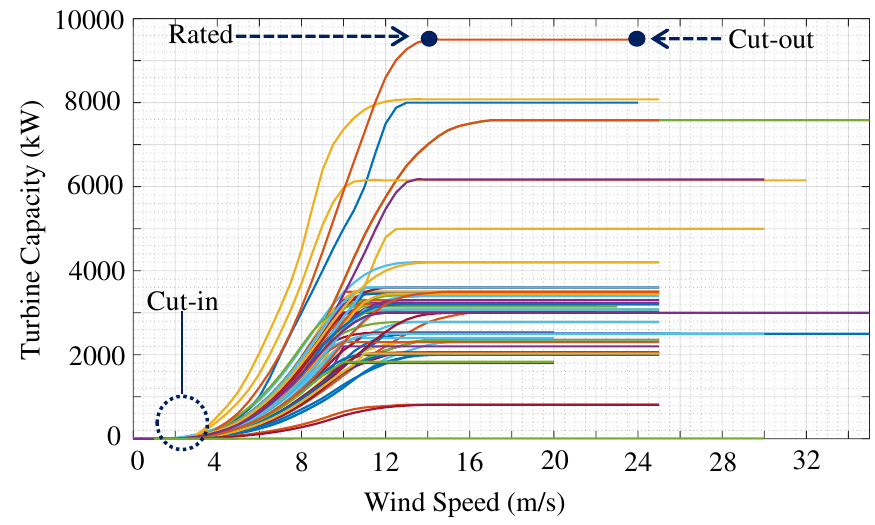}
\caption{Power curves 68 wind turbines with different capacities (10 to 9500 kW)}
\label{Fig:p_curves}
\vspace{-0.2in}
\end{figure}

Turbine data of 68 wind turbines from the most leading turbine manufacturers are used in this study. The data were obtained from the public open energy platform (OEP) \cite{WTDB2019}. Wind turbines data consist of nominal power, rotor diameter, swept area, and power curve values (i.e. wind speed vs. power output). Fig.~\ref{Fig:p_curves} shows the power curves of all turbines with diverse nominal power, cut-in and cut-out values. the International Electrotechnical Commission (IEC) \cite{international2007wind} has established widely accepted standards for wind turbines including the determination of the measured power curve. Commonly, the power curve of the wind turbine is represented using averaged wind speed with intervals of 0.5 m/s \cite{astolfi2018wind}. In this study, the the wind speed intervals (bins) of all wind turbines power curves were limited to 0.5 m/s.

\subsection{Wind Energy Conversion} \label{Sec:EW_Conversion}
The utilized 68 wind turbines operate at different hub-heights which the wind speed data for these hub-height are mostly not available. The wind speed observations of hub-heights 20 m and 26 m above ground for M2 and M4 towers were selected wind turbine power analysis. The power low wind profile (Hellmann exponential law) was used to extrapolate the wind speed observations to higher hub-heights which expressed by:
\begin{equation}
 v = v_0 \bigg( \frac{z}{z_0} \bigg)^\alpha
\label{Eqn:eqn1*}
\end{equation}
\noindent where $v$ is the wind speed at height $z$, $v_0$ is the measured speed where the height $z_0$ is known, and $\alpha$ is the power law exponent (shear exponent or Hellman exponent) describing the terrain topology and varies with atmosphere stability (temperature changes) which commonly set to 1/7, or 0.143 for open land \cite{hossain2018application}. However, in this study, the exponent $\alpha$ parameter is not of concern as much as the stability of wind speed itself for charging EVs.  Therefore, the $\alpha$=0.143 was chosen for this assessment. 


The instantaneous output power of a wind turbine $P_w(t)$ as a function of the wind speed $v$ at turbine hub height $z$, given the power curve $P_c$ is describes as,
\begin{equation}
  P_w(t) =
    \begin{cases}
      0,     & \text{if $v \leq v_{cut\_in}$}\\
      P_c,   & \text{if $v_{cut\_in} < v < v_r$}\\
      P_r,   & \text{if $v_r \leq v < v_{cut\_out}$}\\
      0,     & \text{if $v \geq v_{cut\_out}$}
    \end{cases}   
\label{Eqn:eqn2*}
\end{equation}
\noindent where $v_{cut\_in}$, $v_{cut\_out}$, and $v_r$ are, respectively, the cut-in, cut-out , and nominal (rated) speed of the wind turbine (see Fig.~\ref{Fig:p_curves}). $P_r$ represents the turbine output power at rated wind speed $v_r$. However, the actual wind power output is usually more than the theoretically calculated power as in (\ref{Eqn:eqn2*}), which may produce an estimation error of about 3\% . This is because of the gained momentum by rotating turbine blades sustains the continual rotation when sudden dicrease in wind speed occurs with no major change in the angular wind direction \cite{abolude2018assessment}.

The output power $P(t)$ quantifies the variable turbine power at each time point $t$ (in minutes). However, the charging time of each EV baterry depends on the stability of the supplied power, the battery capacity $B_c$, the battery state-of-charge $SoC$ and the charging point specifications. In order for the EV battery to be fully charged, the charging time $t_{charge}$ can be obtained by \cite{dominguez2019design},
\begin{equation}
 t_{charge} =  \frac{B_c \times (1-SoC)}{P_{charger}}
\label{Eqn:eqn3*}
\end{equation}
Fast charging is a technology that allows higher currents to be delivered to the 
battery before it reaches its peak voltage (usually 80\% $SoC$). Each EV model has 
its predefined parameters such as, $B_c$ and $t_{charge}$ for different $P_{charger}$ capacities. In this study, we consider the $t_{charge}$ value is known, and based on this, the wind power analysis is carried out. 
Due to the existed uncertainty of wind energy over time, the isolated EV charging station is restricted to the stability of the supplied piecewise energy. As shown in Fig.~\ref{Fig:flow_chart} and described in Alg.~\ref{Algo:alg1}, the wind energy fed to the EV station is subject to several constraints: 
\begin{itemize}
\item[-] The instantaneous power contained in each $t_{charge}$ interval must ensure the stability of power supply to EVs with minimum variability, i.e, the standard deviation $\sigma({P_w(t_{charge})})$ must not exceed 0.1 kW. The interval-based average output power is selected such as,

\begin{equation}
\begin{aligned}
 &\overline{P_w} =  \frac{\sum_{t} P_w(t_{charge})}{t_{charge}}, \\ 
 &\text{subject to},\quad  \sigma({P_w(t_{charge})})>0.1\,kW
\label{Eqn:eqn4*}
\end{aligned}
\end{equation}
\item[-] The total wind energy per unit of time (i.e. one day) is calculated from the intervals that show power stability and do not intersect (overlap) in time. Let $P_ov$ to be the instantaneous power intervals of duration $P_{charge}$ distributed  over an entire day time. These intervals are found by using an overlap sliding windows which the overlapped intervals are required to be omitted prior the calculation of total energy for the entire day. The non-overlap total energy is found by,
\begin{equation}
 E_w =  \sum_{i=1}^{N} \overline{P_w^i} \times \frac{t_{charge}}{60}
\label{Eqn:eqn5*}
\end{equation} 
\noindent where $N$ is the total non-overlap intervals (lines 8-10 in Alg.~\ref{Algo:alg1}) within a specific period of time that satisfies the condition $(P_{ov1} \cap P_{ov2} = \phi)$ for each consecutive intervals $P_{ov1}$ and $P_{ov2}$ . 
\item[-] The sum of wind energy generated $E_w^i$ within a $t_{charge}$ time interval is more than the EV charger energy per the smallest time unit (i.e. one minute) for partial charge, or more than the EV charger energy per $t_{charge}$ time for fully charge (80\%).
\end{itemize}

\begin{algorithm}[!t]
\noindent\textbf{Inputs}:\hspace*{2.5mm} $P_w$,\hspace*{6.5mm}The instantaneous wind power;\\ 
\hspace*{14mm} $t_{charge}$, EV battery time to charge;\\
\hspace*{14mm} $t_{ov}$, \hspace*{6mm}Overlap time of intervals;\\
\hspace*{13.5mm} $P_{charge}$, EV charger energy;\\
\noindent\textbf{Outputs}: $E_{nov}$, \hspace*{4mm}Constant supplied energy;\\
\hspace*{13mm} $EV_{no}$, \hspace*{3.5mm}Number of EVs
\vspace{-0.1in}

\noindent\hrulefill
\begin{algorithmic}[1]
\Statex{\textcolor{blue}{/* Calculate the number of overlap intervals */}}
\State {$t_{no}=(P_w.length-t_{charge})/(t_{charge}-t_{ov})$}
\For {$i := 2$ \textbf{to} $t_{no}$ \textbf{step} ($t_{charge}-t_{ov}$)}
\State {$P_{ov1}=P_w[i-1:t_{charge}-1]$}
\State {$P_{ov2}=P_w[i:t_{charge}]$}
\Statex{\textcolor{blue}{/* Check the stability of wind power */}}
\If {std$(P_{ov1})>$0.1 \textbf{or} std$(P_{ov2})>$0.1}
\State {continue}
\EndIf
\Statex{\textcolor{blue}{/* Find non-overlap intervals */}}
\If {$(P_{ov1} \cap P_{ov2} = \phi)$}
\State {$P_{nov}[i]=P_{ov2}$}
\EndIf
\EndFor
\Statex{\textcolor{blue}{/* Calculate the total energy */}}
\State {$E_{nov}=(t_{charge}/60)\sum {P_{nov}[i]}$}
\Statex{\textcolor{blue}{/* Calculate number of EVs */}}
\State {$EV_{no}=({E_{nov}}/{P_{charger}})$}
\end{algorithmic}
\caption{: EV wind power supply}
\label{Algo:alg1}
\end{algorithm}

\subsection{Electric Vehicle Sizing}
In this work, the considered EV fast-charging station consists of renewable sources (wind turbines) and DC power chargers to charge the EVs' batteries. The defined criteria and constraints in Sec. \ref{Sec:EW_Conversion} for the wind-EV charging system are generalized on a specific EV model. For the purposes of this study, we select the Tesla model 3 standard range plus EV, or any other EV with similar specifications. Table \ref{Table:table1} summarizes the relevant characteristics of Tesla model 3+ EV. 

\begin{table}[!t]
\caption{Relevant Tesla model 3 standard range plus specifications}
\vspace{0.15 cm}
\label{Table:table1}
\centering
\resizebox{0.6\textwidth}{!}{
\renewcommand{\arraystretch}{1.15}
\setlength{\tabcolsep}{7pt}
\begin{tabular}{llcccc}
	\hlinewd{0.8pt}
	Parameter & Description or value \\
	\hlinewd{0.8pt}
	Battery pack capacity & 50 kW \\
	Driving range & between 225 -- 465 km\\
	Fast charging range & from 10 -- 80\%\\
	Charging point & Supercharger v3 (250 kW DC)	\\
	Charging point max power & 170 kW\\
	Charging point avg power & 100 kW\\
	DC charging time & 21 minutes\\
	Real energy consumption & between 10.2 -- 21.1 kWh/100 km\\
	\hlinewd{0.8pt}
\end{tabular}}
\vspace{-0.1in}
\end{table}

\vspace{-0.15in}
\section{Experimental Results}
\subsection{Wind Speed Data Averaging}
Three case scenarios of wind speed averaging were used to typify possible variations of  wind power stability by using one, two, and three minutes. For each case, we consider the missing data intervals (for M4 tower) during the averaging process, in which no time interval includes observations from two data segments with missing wind speed values. Fig.~\ref{Fig:missings} shows the monthly percentage of missing data for M4 tower where no missing for M2 tower. These missing data are not scattered over the time but sustain for long periods, sometimes days, and not practical to perform data imputations to fill the data gaps. For this, the power analysis in the following subsections was performed in daily basis. 

Fig.~{\ref{Fig:boxplots} shows the boxplots of entire one year data of M2 and M4 using the three averaging scenarios. Obviously, M2 data have lower annual wind speed average with less variations compared to that of M4. The three averaging intervals also shows almost identical statistics with mean $\pm$ standard deviation of 3.7002$\pm$2.9095, 3.7002$\pm$2.8803, and 3.7002$\pm$2.8616 for one, two, and three minutes intervals. Similarly, for M4 data, the figure shows higher wind averages of 4.4040$\pm$3.5194, 4.4031$\pm$3.4893, and 4.4023$\pm$3.4702, which are almost similar for the different averaging intervals. The stationary analysis of wind averaging also investigated using the obtained wind angular directions and Weibull distribution parameters as shown in Fig~\ref{Fig:rose_dist}. Wind directions of M2 tower as in Fig~\ref{Fig:rose_dist}(a, b, and c) show new patterns with different averaging intervals which could be explained as extreme variations in the wind directions within small time intervals of one to 3 minutes. in contrary, wind directions of M4 tower as in Fig~\ref{Fig:rose_dist}(g, h, and i) are stable over the three time intervals. The wind histograms and Weibull distribution parameters show slight changes when varying the averaging intervals as detailed in Fig~\ref{Fig:rose_dist} caption. These results do not provide an evidence on which averaging interval is more suitable for wind energy harvesting considering stability of power output. Since wind time series data is highly non-stationary, the dynamic analysis of statistical parameters over shorter periods of time may reveal more underlying behaviors of wind changes. This will be provided in the following sections when moving window power analysis is performed over the three averaging scenarios.

\begin{figure}[!t]
	\centering
	\includegraphics[width=0.55\linewidth,keepaspectratio]{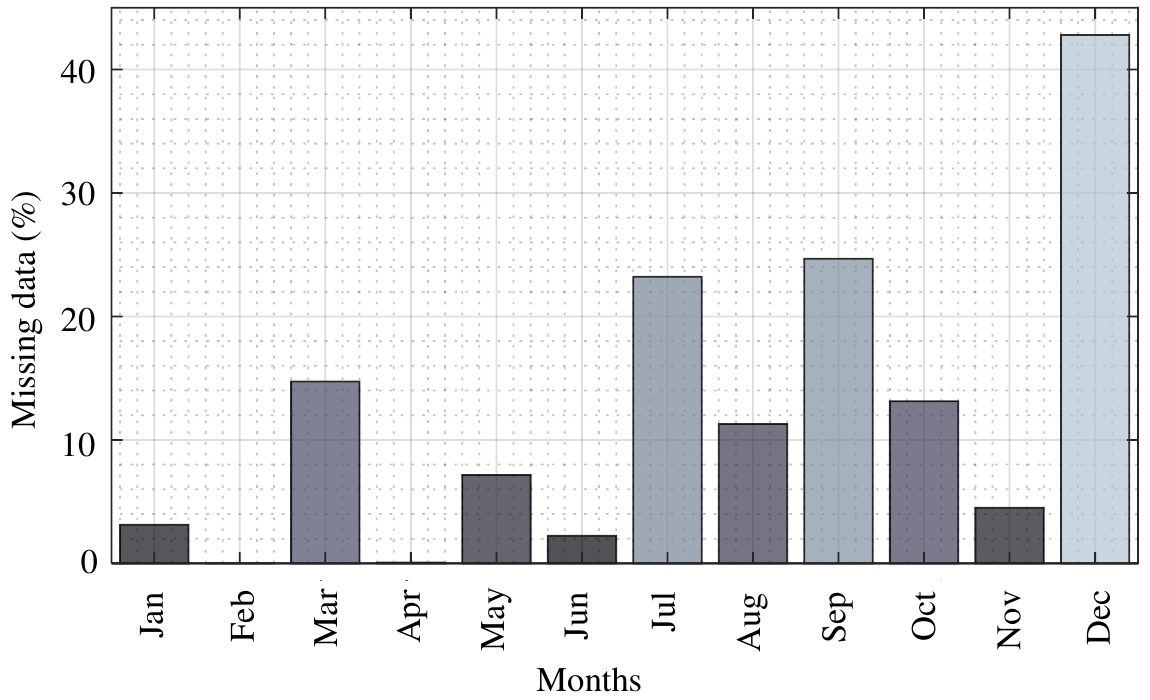}
\vspace{-0.25in}
\caption{Percentage of missing M4-Tower data for each month.}
\label{Fig:missings}
\vspace{-0.2in}
\end{figure}

\begin{figure*}[!t]
	\centering
	\includegraphics[width=18cm,height=10cm,keepaspectratio]{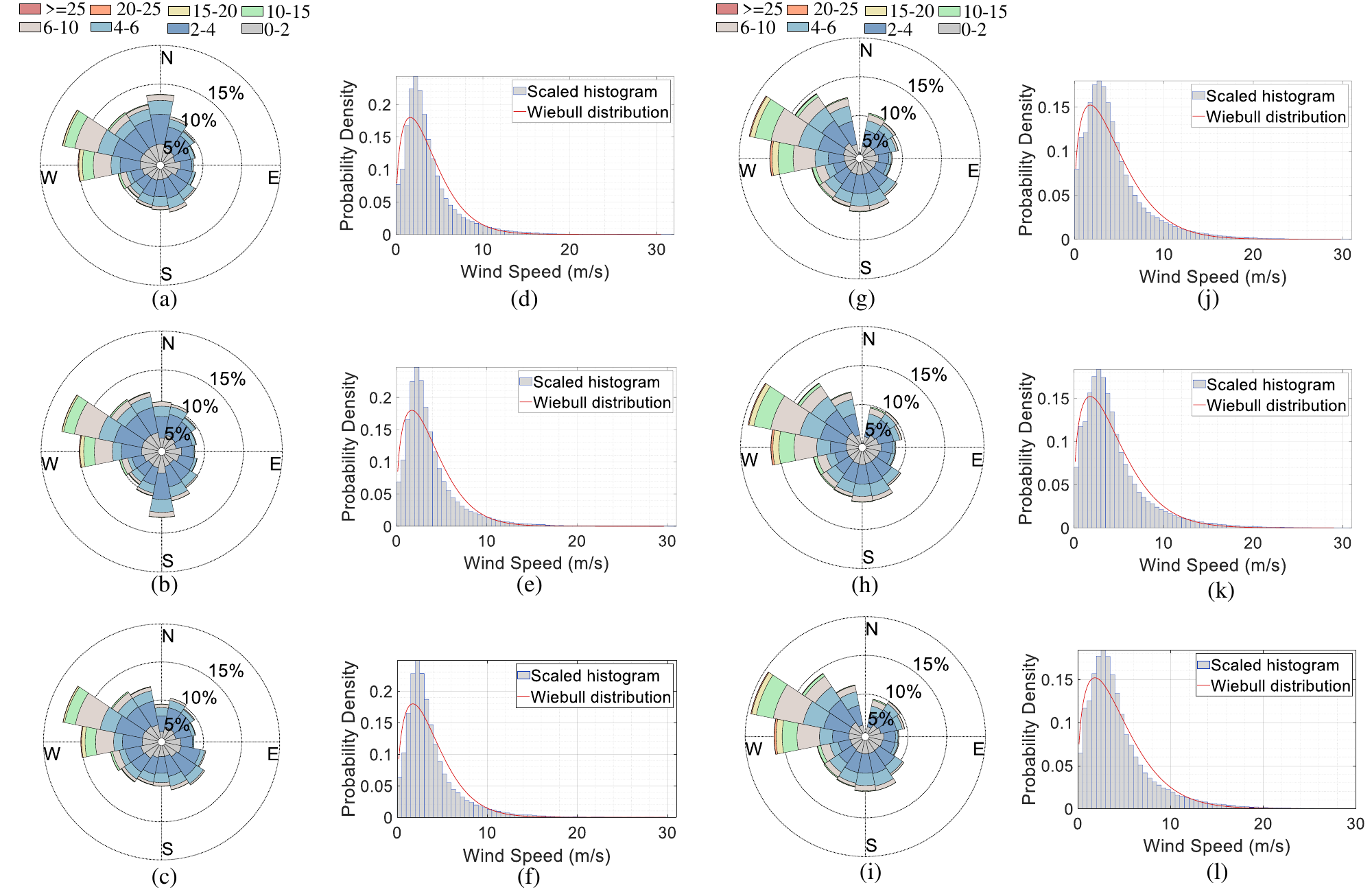}
\vspace{-0.1in}
\caption{Windrose plots and wind speed distributions for M2 (a, b, c, d, e, and f) and M4 (g, h, i, j, k, and l) when averaged over one (a, d, j, and i), two (b, e, h, and k), and three (c, f, i, and l) minutes.(d) Wind speed distribution of M2 averaged over one minute and fitted on Weibull distribution of scale=4.0849 and shape=1.3948. (e) Two minutes averaged wind speed distribution of M2 tower data with Weibull scale=4.0930 and shape=1.410. (f) M2 tower wind speed data averaged over three minutes interval with Weibull scale=4.0986 and shape=1.421. (j) M4 tower wind speed of one minute average with Weibull distribution scale=4.8212 and shape=1.3444. (k) M4 wind speed data when averaging over two minute with Weibull scale=4.8299 and shape=1.3576. (l) Wind speed distribution of M4 tower averaged over three minutes with Weibull distribution scale=4.8357 and shape=1.3669.}
\label{Fig:rose_dist}
\vspace{-0.1in}
\end{figure*} 

\begin{figure}[!t]
	\centering
	\includegraphics[width=0.55\linewidth,keepaspectratio]{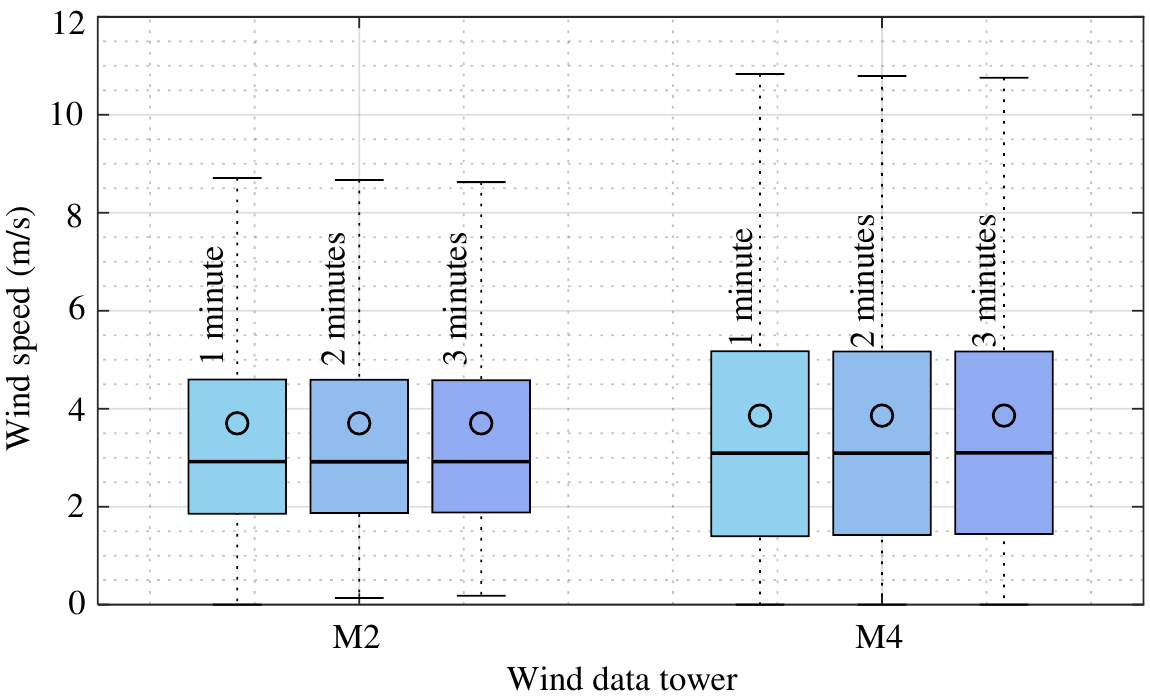}
\vspace{-0.25in}
\caption{Box plot of the M2 and M4 towers wind speed (without ouliers) grouped using the averaging interval, one, two and three minutes.}
\label{Fig:boxplots}
\vspace{-0.3in}
\end{figure}

\subsection{Wind Turbine Selection}
Most wind turbines in the market are of variable speed and produce variable frequency power, which require AC-DC-AC converters to rectify the output power and stabilize the frequency. Variable speed wind turbines are designed to achieve maximum aerodynamic efficiency to make use of wide range of wind speeds. In this study, we perform benchmark performance analysis of the 68 wind turbines to choose the turbines with more stable power outputs. Table~\ref{Table:table1} summarizes the selected nine turbines and the corresponding total $t_{charge}$ intervals percentage when using different averaging intervals for both M2 and M4 towers data. The optimal selection of wind turbines may depend on other indicators, such as the total energy production, the cost of produced energy, reliability, and installation cost. However, the main concern in this study is the availability of stable output power; hence, the turbines with more near-constant power intervals are selected for EV charging analysis. From Table~\ref{Table:table1}, the selected turbines have nominal power in the range of 2200--3500 kW with hub heights from 99.5--137 m. 

\begin{table}[!t]
\caption{Selected wind turbines output power stability (\%) using M2 and M4 wind speed data with different averaging intervals.}
\vspace{0.15 cm}
\label{Table:table2}
\centering
\resizebox{0.7\textwidth}{!}{
\renewcommand{\arraystretch}{1.15}
\setlength{\tabcolsep}{3pt}
\begin{tabular}{llllllccccccc}
	\hlinewd{0.8pt}
&&&&&Power&\multicolumn{3}{c}{M4-Tower data} & & \multicolumn{3}{c}{M2-Tower data}\\
ID	    &height	   &\text{cut\_in}&rated&\text{cut\_out}& (kW)	&1Min	&2Min	&3Min	&	&1Min	&2Min	&3Min\\
\hlinewd{0.8pt}
no16	&134	   &3&10&20&3300&9.41	&14.34	&18.35&	&5.22	&8.28	&9.78\\
no17	&114	   &3&10&20&3000&9.22	&14.44	&17.74&	&5.22	&7.74	&9.64\\
no44	&119	   &3&10&22&3000&12.61	&17.69	&21.26&	&6.82	&9.12	&10.49\\
no67	&129	   &3&10&23&3150&13.41	&17.94	&22.20&	&7.20	&9.41	&10.87\\
no73	&99.5      &3&10&25&2300&14.96	&18.73	&22.20&	&6.92	&8.97	&10.21\\
no94	&139	   &3&10&22&3200&12.42	&17.05	&21.45&	&7.10	&9.41	&11.01\\
no95	&136	   &3&10&22&3000&12.33	&17.00	&21.31&	&7.10	&9.46	&10.91\\
no124	&137	   &3&10&25&3500&16.04	&20.21	&23.76&	&7.86	&9.96	&11.34\\
no128	&99	       &3&10&25&2200&14.96	&18.73	&22.20&	&6.87	&8.97	&10.21\\
\hlinewd{0.8pt}
\end{tabular}}
\vspace{-0.1in}
\end{table}

\subsection{Electric Vehicle Sizing}
The wind powered EV charging station is strongly dependent on the availability of constant power supply from wind turbines which limits the station to provide smart charging rather than immediate charging scenario. A grid power compensation could partially solve this issue and provide enough power to charge the EVs when turbines fail to satisfy the EVs power demand. In this study, the direct wind-EV charging using DC fast charging technology is preferable due to, the significant reduction in power conversion stages, avoiding the use of power storage systems, and the excess wind power could be ejected to the utility grid. The (re-)schedule of charging event is triggered whenever the charging system predicts a stable wind energy falls within a user-defined EV charging specifications (energy volume and charging duration). All time intervals (holding time of EV for charging) were statistically analyzed filtering the intervals that are not able to show stable or enough power within the EV standards. The remaining time charging intervals were used to find the cumulative daily wind energy and the estimated number of EVs that station could handle. 

Fig.~\ref{Fig:tot_evs} summarizes the average monthly number of EVs calculated from each individual wind turbine power output over one, two, and three minutes wind speed averaging for M2 and M4 data sets. For all scenarios, turbine no124 shows the best performance from the selectec nine turbines followed by turbine no44. Even though the wind speed data variations and power curves' sampling is the same for all turbines, instantaneously mapping the wind speed to wind power using the power curve induces more stability in the output wind power which strongly depends on the shape of the power curve itself.

Also, comparison results of using different wind speed averaging intervals reveals that, as the averaging interval (in minutes) increases, the more stable the wind power output which verifies the blades momentum effects. Comparing the results of one minute averaging (Fig.\ref{Fig:tot_evs}(a and c)) with the two minutes averaging (Fig.\ref{Fig:tot_evs}(b and d)), it shows that there is a significant increase in power stability which leads to more number of EVs the station can serve in monthly basis. For turbine no16 (January) as an example, the two minutes averaging intervals show about 47\% and 45\% increase in total EVs for M2 and M4 data sets respectively. Similarly, the three minutes compared to one minute averaging interval show superior improvement in the system, producing almost 81\% and 91\% increase in system monthly coverage of EVs.  

Furthermore, Fig~\ref{Fig:tot_evs} illustrates the comparison results of using two different wind speed data sets (M2-2018 and M4-2014). According to the results, it can be seen that M4 data (Fig~\ref{Fig:tot_evs}(d, c, and f)) provides more wid turbine power output and, therefore, more number of EVs to charge. However, the performance is not stable throughout the months of the year. M2 data set shows best performance during Jan, Feb, Mar, Apr, Nov, and Dec, with low performance during summer months. On the other hand, M4 data also follows the same conclusion of M2, except for Dec month where almost 12 days of data were missing which were excluded from the calculations. 

\begin{figure*}[!t]
	\centering
	\includegraphics[width=18cm,height=10cm,keepaspectratio]{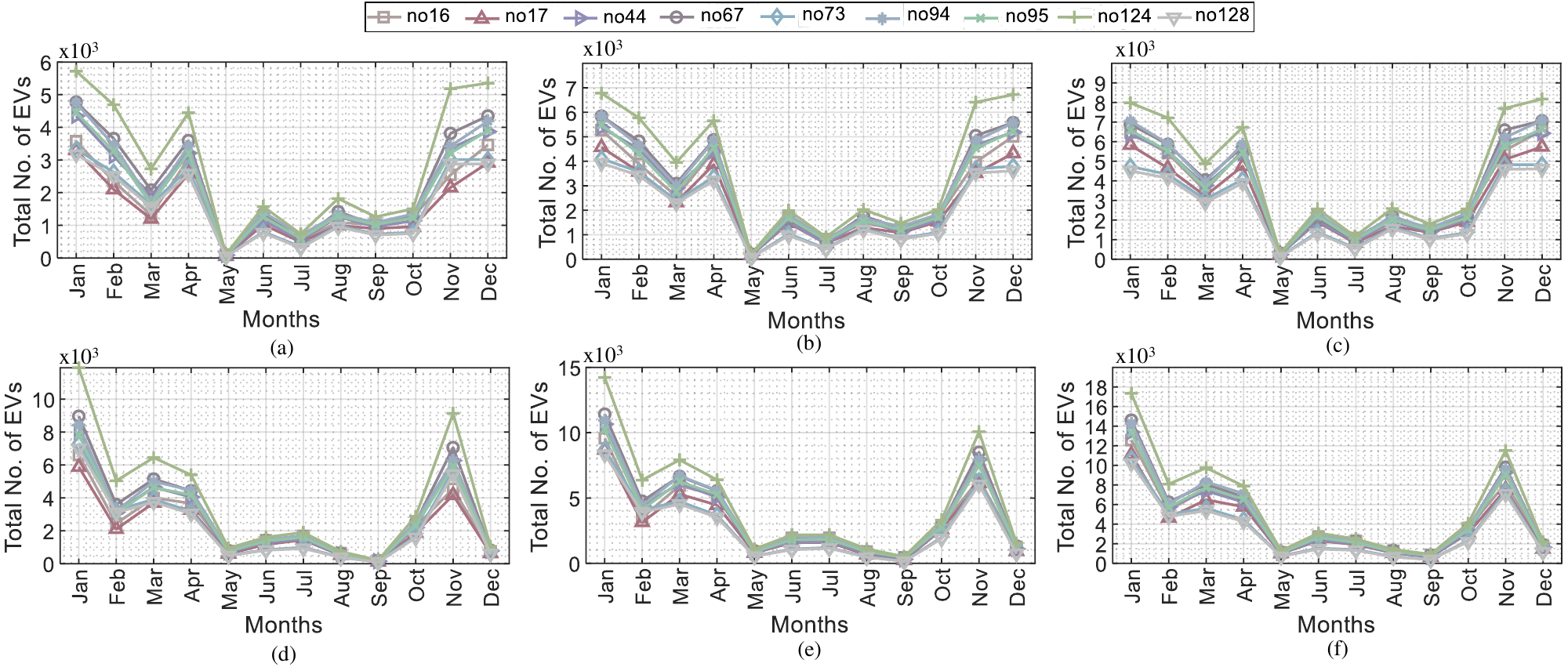}
\vspace{-0.1in}
\caption{Selected turbines and total number of EVs calculated over different averaging intervals for M2 (a,b, and c)and M4 (d, e, and f) data. (a and d) Show the monthly estimated total EVs when using one minute wind speed averaging. (b and d) Number of EVs using two minutes averaging intervals. (c and f) The three minute averaging results of total EVs.}
\label{Fig:tot_evs}
\end{figure*} 

\begin{figure*}[!h]
	\centering
	\includegraphics[width=18cm,height=10cm,keepaspectratio]{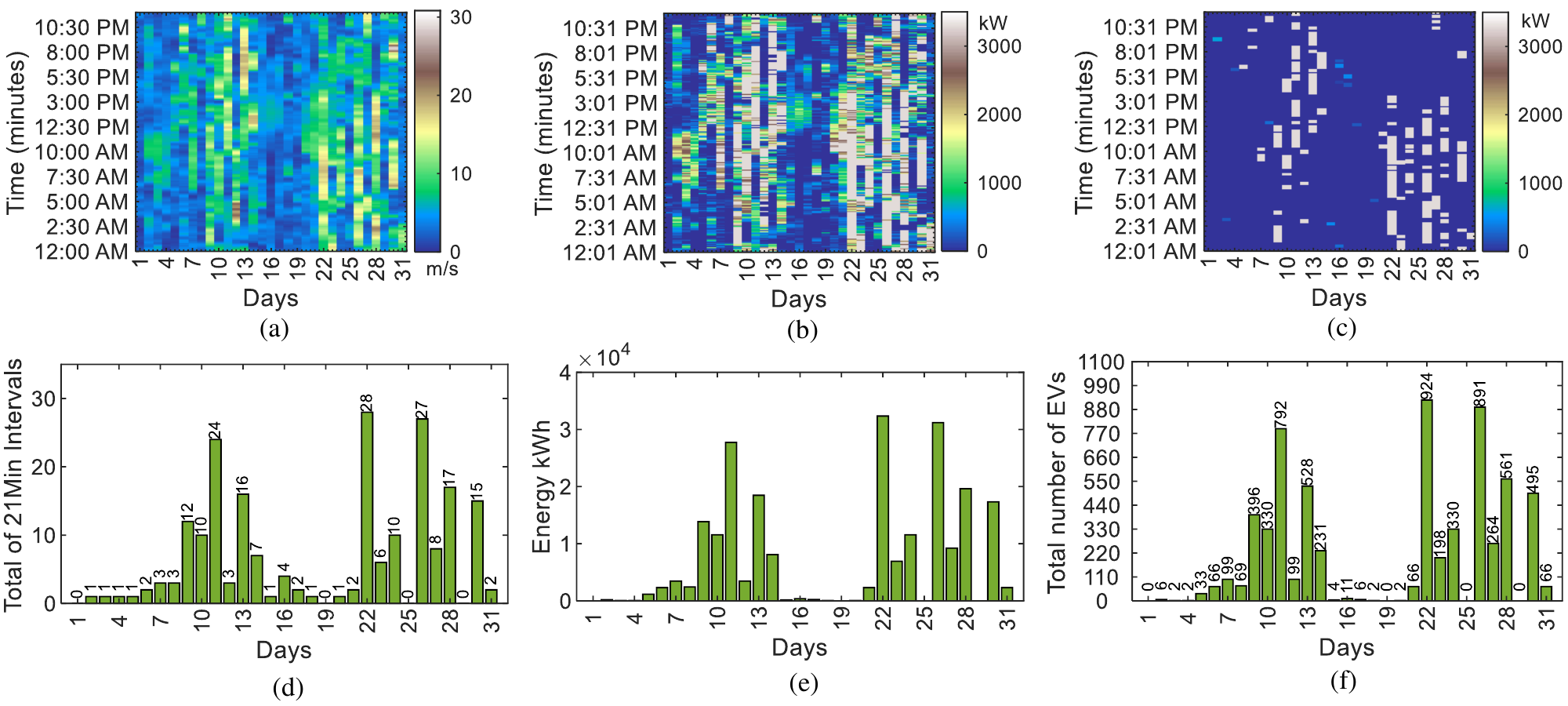}
\vspace{-0.1in}
\caption{Wind energy EV charging conversion results for M2 tower data during January-2018 for selected turbine No.16. (a) wind speed (average $\pm$ standard deviation is 5.6898 $\pm$ 4.1389 m/s) heat-map rearranged to show daily 21-minutes intervals. (b) The corresponding wind turbine power output for each 21-minute interval. (c) The usable interval for EV charging. (d) Total number of 21-minute intervals per day. (e) Total available daily energy for full-EV charging. (f) Total number of EVs per day.}
\label{Fig:ev_m2}
\vspace{-0.2in}
\end{figure*} 

Fig.~{\ref{Fig:ev_m2} and Fig.~{\ref{Fig:ev_m4} show examples of direct wind energy EV charging conversion procedure of one month for both M2 (January, 2018) and M4 (January, 2014) data respectively. The wind speed data was averaged using three minutes intervals as concluded from Fig~\ref{Fig:tot_evs}. Fig.~{\ref{Fig:ev_m2}(a) and Fig.~{\ref{Fig:ev_m4}(a) show the heat-maps of interval-based wind speed progression (m/s) rearranges in daily vectors to observe the correlation between wind speed observations and wind turbine output power. For each pre-defined charging time slot ($t_{charge}=21$ minutes), the algorithm converts the wind speeds to turbine power outputs using overlapped sliding windows procedure (Fig.~{\ref{Fig:ev_m2}(b) and Fig.~{\ref{Fig:ev_m4}(b)). It can be concluded that the regions in the wind speed heat-maps that show no changes in the color (wind speed) are reflected as stable power outputs in the corresponding power heat-maps. The filtered wind power time slots ($t_{charge}=21$ minutes) are shown in Fig.~{\ref{Fig:ev_m2}(c) and Fig.~{\ref{Fig:ev_m4}(c). We can clearly see that the power time slots which are constrained for EV charging are scattered throughout the day time with M4 data set show more availability of EV charging time slots. For power time slots available at night time and for better charging system efficiency, prior alerts must be sent to the customers when the power is expected to be available and parking lots availability in the charging station. 

\begin{figure*}[!t]
	\centering
	\includegraphics[width=18cm,height=10cm,keepaspectratio]{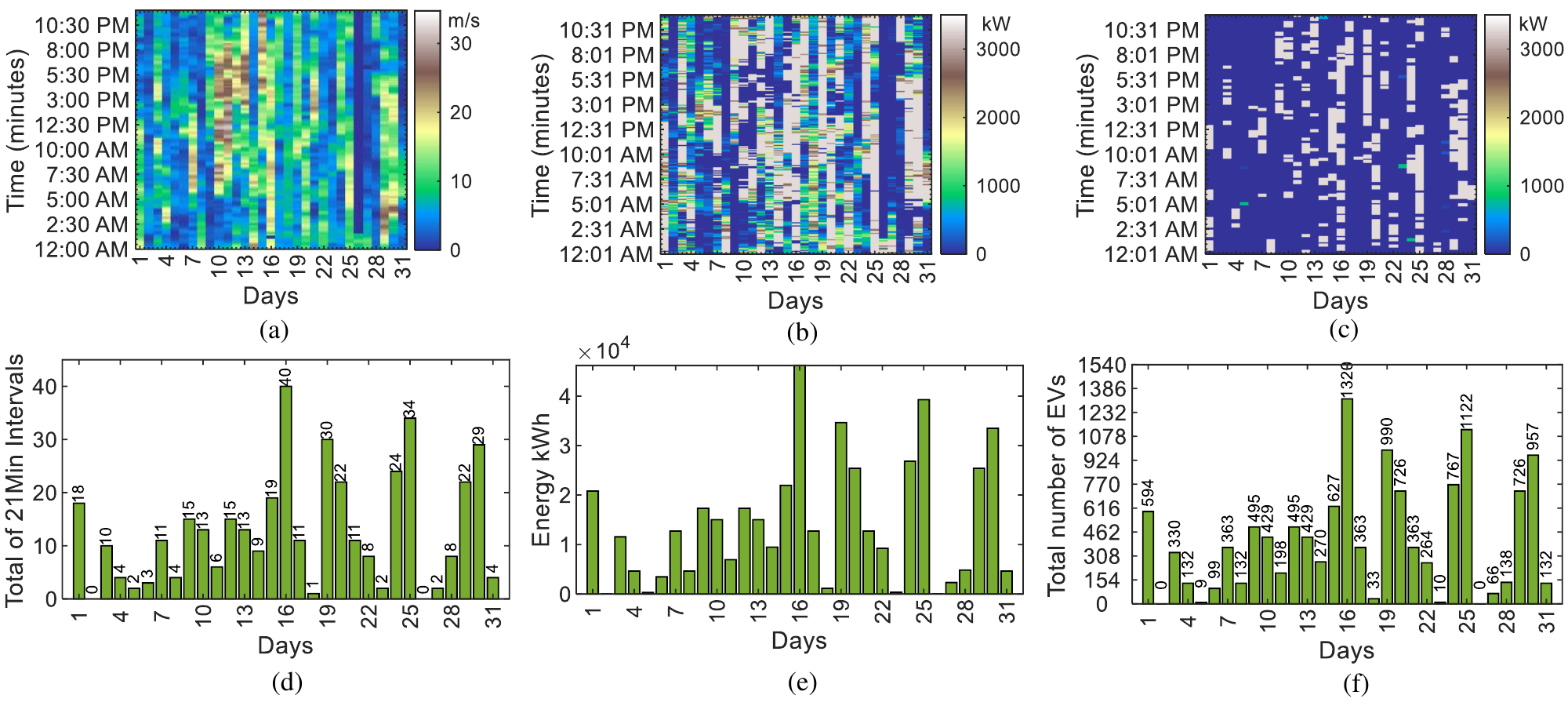}
\vspace{-0.1in}
\caption{Wind energy EV charging conversion results for M4 tower data during January-2014 for selected turbine No.16. (a) wind speed (average $\pm$ standard deviation is 8.3917 $\pm$ 6.2554 m/s) heat-map rearranged to show daily 21-minutes intervals. (b) The corresponding wind turbine power output for each 21-minute interval. (c) The usable interval for EV charging. (d) Total number of 21-minute intervals per day. (e) Total available daily energy for full-EV charging. (f) Total number of EVs per day.}
\label{Fig:ev_m4}
\vspace{-0.2in}
\end{figure*} 

The conversion of wind power to energy can be seen in Fig.~{\ref{Fig:ev_m2}(d and e) and Fig.~{\ref{Fig:ev_m4}(d and e). Bot M2 and M4 data sets show different patterns of daily energy availability, with minimal EV charging energy in some days, especially for M2 data. We suggest to add another renewable energy source, connect to the grid, or both scenarios to tackle this energy uncertainty. Though, the wind energy excess power can be injected to the utility grid during the windy days which will compensate the system energy cost provided to the customers. As shown in Fig.~{\ref{Fig:ev_m2}(f) and Fig.~{\ref{Fig:ev_m4}(f), only few days with zero wind energy which was harvested from individual wind turbine, were wind energy aggregation from multi-turbine grid could enhance the overall system efficiency. Furthermore, for some days, the wind energy availability for EV charging could be beyond the charging station capacity, which in turns, this energy could be sold to the utility grid and re-used when wind energy is not available rather than using battery storage systems.

\section{Conclusion}
\label{sec:prior}
In this paper, we described a fast EV charging approach, which considers wind turbines as a direct energy source to the charging station. The benefits of this approach can be realized from the reduced dependency on grid utility, battery storage systems, and energy conversion power electronics. We conducted wind speed and power data analysis of high resolution data sets acquired from two different sites within two distanced years. We performed investigations of wind averaging intervals. The results concluded that averaging wind speed over intervals of three minutes show better EV charging performance.

We also conducted a comparison of different wind turbines using large (68 turbines) power curves' data library. Analysis results reveal only nine turbines to show better stable power availability for EV charging. We presented the proposed charging approach by using a standard EV (Tesla model 3 standard range plus) to analytically measure the overall performance of the charging approach. The analysis came to a conclusion, that the introduced charging approach is able to provide convincing results. 

\bibliographystyle{IEEEbib_ori}
\bibliography{Ref-EVDC-TIA-reduced}

\end{document}